\documentclass[fleqn,usenatbib]{mnras}
\usepackage{newtxtext,newtxmath}
\usepackage[T1]{fontenc}
\DeclareRobustCommand{\VAN}[3]{#2}
\let\VANthebibliography\thebibliography
\def\thebibliography{\DeclareRobustCommand{\VAN}[3]{##3}\VANthebibliography}
\usepackage{graphicx}	
\usepackage{amsmath}	
\usepackage{threeparttable}
\usepackage [caption=false]{subfig}
\defcitealias{Hagen_2024b}{H24}
\usepackage{xcolor}

\title[Collapse of the disc in AGN]{Systematic collapse of the accretion disc in AGN confirmed by UV photometry and broad line spectra}

\author[Kang et al.]{Jia-Lai Kang$^{1,2,3}$\thanks{jialai.kang@durham.ac.uk},
Chris Done$^{1,4}$\thanks{chris.done@durham.ac.uk},
Scott Hagen$^{1}$,
Matthew J. Temple$^{1}$, 
John D. Silverman$^{4,5,6,7}$,
\newauthor
Junyao Li$^{8}$,
Teng Liu$^{2,3}$
\\
$^{1}$Centre for Extragalactic Astronomy, Department of Physics, Durham University, South Road, Durham, DH1 3LE, UK\\
$^{2}$Department of Astronomy, University of Science and Technology of China, Hefei 230026, China \\
$^{3}$School of Astronomy and Space Science, University of Science and Technology of China, Hefei 230026, China \\
$^{4}$Kavli Institute for the Physics and Mathematics of the Universe (Kavli IPMU, WPI), UTIAS, Tokyo Institutes for Advanced Study, University of Tokyo, \\ Chiba, 277-8583, Japan\\
$^{5}$Department of Astronomy, School of Science, The University of Tokyo, 7-3-1 Hongo, Bunkyo, Tokyo 113-0033, Japan \\
$^{6}$Center for Data-Driven Discovery, Kavli IPMU (WPI), UTIAS, The University of Tokyo, Kashiwa, Chiba, 277-8583, Japan \\
$^{7}$Center for Astrophysical Science, Department of Physics \& Astronomy, John Hopkins University, Baltimore, MD 21218, USA \\
$^{8}$Department of Astronomy, Univerity of Illinois at Urbana-Champaign, Urbana, IL 61801, USA \\
}

\date{Accepted XXX. Received YYY; in original form ZZZ}

\pubyear{\the\year{}}

\begin{document}
\label{firstpage}
\pagerange{\pageref{firstpage}--\pageref{lastpage}}
\maketitle

\begin{abstract}
{
A recent study on the spectral energy distribution (SED) of AGN combined unobscured X-ray sources from the eROSITA eFEDS Survey with high quality optical imaging from Subaru’s Hyper Suprime-Cam (HSC). The HSC data enabled accurate host galaxy subtraction as well as giving a uniform black hole mass estimator from the stellar mass. 
The resulting stacked optical/X-ray SEDs for black holes at fixed mass show a dramatic transition, where the dominating disc component in bright AGN evaporates into an X-ray hot plasma below $L/L_{\rm Edd}\sim 0.01$. The models fit to these datasets predicted the largest change in SED in the rest frame UV ($< 3000$\,\AA), but this waveband was not included in the original study. Here we use archival $u$-band and UV photometry to extend the SEDs into this range, and confirm the UV is indeed intrinsically faint in AGN below $L/L_{\rm Edd}\sim 0.01$ as predicted. This dramatic drop in UV photo-ionising flux is also seen from its effect on the broad emission lines. We stack the recently released SDSS DR18 optical spectra for this sample, and show that the broad H$\beta$
line disappears along with the UV bright component at $L/L_{\rm Edd}\sim 0.01$. This shows that there is a population of unobscured, X-ray bright, UV faint AGN which lack broad emission lines (true type 2 Seyferts).

}

\end{abstract}

\begin{keywords}
accretion, accretion discs -- black hole physics -- galaxies: active
\end{keywords}

\section{Introduction}

\par Active Galactic Nuclei (AGN) are powered by accretion onto a supermassive black hole (SMBH), producing strong emission over a wide electromagnetic range. The classic ``standard'' model of the accretion flow is an optically thick, geometrically thin disc \citep[SS disc, ][]{Shakura_1973}, where the gravitational energy released at each radius thermalizes locally to produce a (quasi)-blackbody spectrum. 
The disc temperature increases at smaller radii such that the total spectrum is a sum of blackbody components (i.e a multicolour blackbody) \citep[e.g.,][]{Mitsuda_1984}, which typically peaks in the extreme ultraviolet (EUV) for bright AGN \citep[e.g.,][]{Elvis_1994}. This peak is not easily observerd due to interstellar absorption within our Galaxy, though models generally predict a strong blue component, as observed \citep[the ``big blue bump'',][]{Malkan_1982, Sanders_1989, Laor_1989}, as well as prominant emission lines \citep[e.g.,][]{Vanden_2001} produced from photoionisation by the strong EUV disc continuum.

\par However, the accretion flow in AGN is clearly more complicated than a standard SS disc. AGN show ubiquitous X-ray emission \citep[e.g.,][]{Elvis_1994, Lusso_2016}, with a typical power-law continuum extending to a few hundred keV \citep{Walter_1993, Kang_2022}. This cannot originate from the disc. Instead they indicate the presence of an optically thin hot plasma (i.e the corona) near the central black hole, where seed photons from the disc are Compton up-scattered to X-ray energies, producing the power-law continuum \citep{Haardt_1991, Haardt_1993}. 
Below $\sim 2$\,keV, however, the X-ray spectrum often deviates from a single power-law continuum, instead showing an upturn \citep[soft X-ray excess,][]{Boller_1996, Gierlinski_2004, Bianchi_2009}. This can be modelled as partially ionised reflection from the accretion disc \citep[e.g.,][]{Crummy_2006}. However, the upturn generally appears to connect to a downturn in the UV, occurring before its expected peak \citep{Zheng_1997, Telfer_2002, Shull_2012}. Hence, the UV downturn and soft X-ray excess can instead be modelled together by a warm Comptonisation component \citep[i.e warm corona,][]{Magdziarz_1998, Done_2012, Jin_2012, Jin_2017, Petrucci_2018}. This can dominate the bolometric luminosity, and is optically thick, so is likely the disc itself. However its spectrum is clearly not thermalising into a blackbody as in SS disc models.

\par The SS disc models also fail to explain the variability of AGN, which can be 
more than an order of magnitude \citep{Ren_2022}
on time-scales of months to years in the UV/optical band \citep[e.g.,][]{Vanden_2004, MacLeod_2010}, while the typical viscous timescale of a standard disc is thousands of years \citep{Noda_2018, Lawrence_2018}. A potential solution is provided by the X-ray reprocessing model \citep{Clavel_1992, Cackett_2007}, where the variable X-rays produced by the compact hot corona illuminate the disc, driving the UV/optical variations. However, intensive monitoring projects now 
clearly show that the observed optical fluctuations are poorly modelled by
X-ray reprocessing \citep[e.g., the compilation of][]{Edelson_2019}. It is possible that 
X-ray reprocessing is the driver of the observed optical variability if the hot corona geometry is also changing in a complex way with X-ray flux \citep{Kammoun_2023}, but it seems more likely that the 
observed optical variability is indeed intrinsic \citep{Cai_2018, Cai_2020}, unlike the predictions of the standard disc model \citep{Hagen_2023,Hagen_2024}.

\par Intrinsic disc variability is clearly seen in the rare ``changing-look'' or ``changing-state'' AGN \citep[e.g.,][]{LaMassa_2015, Ruan_2016, Yang_2018, Temple_2023}. These objects show the appearance/disappearance of the broad emission lines on a typical timescale of years, often along with strong UV/optical continuum variations of a factor $>$\,10 \citep{MacLeod_2016}. A large fraction of these have correlated IR variability, showing that the UV variability affects the irradiating flux on the torus, so its variability must be due to an intrinsic 
change of the accretion state rather than changing obscuration along our line of sight \citep{Sheng_2017}. The changing SED of the accretion flow is seen directly in detailed studies of individual objects, 
e.g. Mrk 1018 \citep{Noda_2018}, where the
bright UV collapses at Eddington ratio $L/L_{\rm Edd} =\dot{m}\sim 0.01$ \citep{Noda_2018}, similar to the state transitions widely observed in X-ray binaries \citep[see the review by]{Done_2007}. 

\par This can be interpreted as the luminosity at which the standard disc transitions into a hot, geometrically thick and radiatively inefficient accretion flow, \citep[e.g.,][]{Narayan_1995,Yuan_2007}. The SED of low luminosity/accretion rate AGN do appear consistent with this, as they lack a strong UV disc component
\citep{Ho_1999, Laor_2003, Elitzur_2009, Trump_2011}. 
However, these are observationally very difficult to select and investigate due to the
much stronger contamination from the 
host galaxy emission drowning out the AGN, limiting the previous analyses to individual sources or small samples.

\par These limitations were circumvented by \citet{Hagen_2024b} (hereafter \citetalias{Hagen_2024b}).
They used eROSITA X-ray selected AGN from the eFEDS field \citep{Brunner_2022, Liu_2022} to unambiguously identify the AGN, together with high-quality Subaru Hyper Suprime-Cam (HSC) multi-band images \citep{Aihara_2022} allowing for the decomposition of the host galaxy from the AGN emission by \citet{Li_2024}.
The resulting AGN spectra were binned as a function of black hole mass (using the stellar mass as a proxy) and AGN optical luminosity. Taking the single mass bin of $\log M_{\rm BH}=8-8.5$ gives a sequence of SEDs as a function of Eddington ratio, 
$\dot{m} = \dot{M}/\dot{M_{\rm{Edd}}} = L/L_{\rm{Edd}}$. \citetalias{Hagen_2024b} found 
that the resulting SEDs show a dramatic drop of the blue disc continuum for sources with $\dot{m}$ below $\sim 0.01$. Their modelling predicted that the largest change in the SED should be in the UV/EUV, however this waveband was not covered by their data. 

\par In this work we extend these SEDs from \citetalias{Hagen_2024b} to the rest-frame UV using 
SDSS-u, KiDS-u, GALEX-NUV and GALEX-FUV photometry. This confirms the dramatic drop in the 
UV continuum predicted by the models for $\dot{m}\lesssim 0.01$,
indicating a collapse of the optically thick accretion disc emission for low luminosity AGN. 

\par The dramatic drop in UV photoionizing flux predicts a similar drop in the strength of the broad emission lines, as discussed in \citetalias{Hagen_2024b}. Only SDSS DR16 was available at time of their study, which 
did not include many of the lower luminosity AGN in this sample. However, the latest data release of SDSS \citep[DR18, ][]{Almeida_2023} targets
a large number of eFEDS sources, as part of the SDSS-V/eFEDS survey, which now enables a detailed spectral investigation. These stacked spectra show the broad emission lines are 
indeed absent or extremely weak in sources with $\dot{m}\lesssim 0.01$, as predicted by the change in the continuum. 

\par While our new study generally confirms the SED models of \citetalias{Hagen_2024b}, it does also reveal a small discrepancy in the higher luminosity AGN. The disc models based on 
optically thick, warm Comptonisation give a smoothly increasing continuum flux, whereas the UV data have a generic break at $\sim 1000$\,\AA\
as seen in previous work focused on the 
UV bandpass \citep{Zheng_1997, Telfer_2002, Shull_2012, Cai_2023, Cai_2024}. This must be intrinsic rather than connected to dust in either the AGN or the host galaxy as the X-ray selection only included unobscured AGN.
This likely points to the importance of atomic physics in the disc, perhaps due to UV line driven winds \citep{Laor_2014}.

The paper is organised as follows. In section 2 we detail the sample and data selection. In section 3 we show the impact the ancillary UV data have on the SEDs, and in section 4 we show the impact on the broad emission lines. As in \citetalias{Hagen_2024b} we assume a standard cosmology, from the Planck\,2018 results \citep{Planck_2020}.

\section{Sample and Data}

\subsection{An eFEDS-HSC AGN sample}
\par We start from the subsample in \citetalias{Hagen_2024b}, composed of 1305 AGN with black hole masses of log($M_{\rm BH}$/$M_{\sun}$) between 8.0 and 8.5. In brief, this  consists of low-absorption (intrinsic $N_{\rm H} < 10^{22}\, \rm cm^{-2} $) AGN from eFEDS \citep{Liu_2022}, with high-quality Subaru Hyper Suprime-Cam (HSC) multi-band images \citep{Li_2024}, which allow effective AGN-host decomposition \citep{Ishino_2020, Li_2021a}. The black hole masses are estimated via the $M_{\rm{BH}} - M_{\rm{stellar}}$ relationship \citep{Ding_2020, Li_2021b}, where the stellar mass is measured from the HSC images \citep{Li_2024}.
Compared with previous optically selected samples, this sample contains sources spanning a wide luminosity range (over three orders of magnitude in the optical) with similar black hole masses, and particularly a large number of low-luminosity sources (see Figure\,3 in \citetalias{Hagen_2024b}). This provides a unique opportunity to investigate how the SED evolves with accretion rate, within the range of $ 0.01 \lesssim \dot{m} \lesssim 1.0$. We now extend these into the UV, in order to critically test the predicted collapse of the optically thick disc emission.

\begin{center}
\begin{table*}
 \footnotesize
	\centering
	\caption{Stacked luminosities for the eight bins of $\nu L_{3500}$. 
    }
    \label{tab:result}
 \renewcommand\tabcolsep{4pt}
\begin{tabular}{ccccccccc} 
\hline
log $\nu L_{3500}$ & [42.1, 42.5] & [42.5, 42.9] & [42.9, 43.3] & [43.3, 43.7] & [43.7, 44.1] & [44.1, 44.5] & [44.5, 44.9] & [44.9, 45.3] \\
log $\dot{m}$ & -1.93 & -2.03 & -2.04 & -1.93 & -1.80 & -1.40 & -0.98 & -0.61 \\
Source number & 36 & 146 & 287 & 333 & 247 & 164 & 76 & 16 \\
Mean redshift & 0.36 & 0.46 & 0.51 & 0.56 & 0.57 & 0.60 & 0.66 & 0.71 \\
Mean log ($N_{\rm H, int}$) & 20.96 & 20.94 & 20.89 & 20.74 & 20.61 & 20.43 & 20.41 & 20.35 \\
\hline
FUV detected / covered & 4 / 24  & 10 / 88  & 43 / 202  & 73 / 239  & 96 / 181  & 105 / 134  & 52 / 62  & 11 / 11  \\
FUV $\lambda$ (\AA) & - & 1045  & 1013  & 983  & 977  & 955  & 925  & 895  \\
FUV log $\nu L_{\nu}$ & - & 41.83 $\pm$ 0.10  & 43.16 $\pm$ 0.10  & 43.34 $\pm$ 0.08  & 43.88 $\pm$ 0.08  & 44.30 $\pm$ 0.02  & 44.65 $\pm$ 0.13  & 45.14 $\pm$ 0.06  \\
\hline
NUV detected / covered & 5 / 24  & 18 / 88  & 64 / 202  & 107 / 239  & 127 / 181  & 121 / 134  & 56 / 62  & 11 / 11  \\
NUV $\lambda$ (\AA) & - & 1554  & 1507  & 1463  & 1449  & 1415  & 1376  & 1333  \\
NUV log $\nu L_{\nu}$ & - & - & 42.88 $\pm$ 0.07  & 43.46 $\pm$ 0.04  & 43.96 $\pm$ 0.02  & 44.50 $\pm$ 0.03  & 44.88 $\pm$ 0.05  & 45.31 $\pm$ 0.06  \\
\hline
 Both detected / covered & 2 / 24 & 9 / 88 & 40 / 202 & 69 / 239 & 89 / 181 & 103 / 134 & 50 / 62 & 11 / 11 \\
 log $\nu L_{1200}$ & - & - & 43.06 $\pm$ 0.09  & 43.40 $\pm$ 0.12  & 43.93 $\pm$ 0.09  & 44.42 $\pm$ 0.04  & 44.80 $\pm$ 0.05  & 45.27 $\pm$ 0.02  \\
\hline
SDSS-$u$ detected / covered & 3 / 26  & 36 / 102  & 86 / 221  & 154 / 277  & 172 / 226  & 150 / 160  & 74 / 76  & 15 / 15  \\
SDSS-$u$ $ \lambda$ (\AA) & - & 2500  & 2400  & 2290  & 2255  & 2222  & 2142  & 2083  \\
SDSS-$u$ log $\nu L_{\nu}$ & - & 43.00 $\pm$ 0.05  & 43.30 $\pm$ 0.05  & 43.70 $\pm$ 0.02  & 43.97 $\pm$ 0.02  & 44.42 $\pm$ 0.03  & 44.78 $\pm$ 0.04  & 45.18 $\pm$ 0.02  \\
\hline
KiDS-$u$ detected / covered & 6 / 11  & 27 / 49  & 76 / 118  & 133 / 165  & 117 / 118  & 85 / 85  & 34 / 34  & 9 / 9  \\
KiDS-$u$ $ \lambda$ (\AA) & - & 2419  & 2362  & 2290  & 2290  & 2205  & 2127  & 2127  \\
KiDS-$u$ log $\nu L_{\nu}$ & - & 42.83 $\pm$ 0.04  & 43.06 $\pm$ 0.02  & 43.48 $\pm$ 0.02  & 43.87 $\pm$ 0.02  & 44.34 $\pm$ 0.03  & 44.68 $\pm$ 0.03  & 45.08 $\pm$ 0.03  \\
\hline
Number of SDSS Spectra & 7 & 26 & 47 & 88 & 87 & 96 & 49 & 12 \\
\hline
\end{tabular}
\begin{tablenotes}
\item{*}{All $\nu L_{\nu}$ values are stacked luminosities with units of $\rm erg\,s^{-1}$ and all $\lambda$ values are mean effective rest-frame wavelengths of each band. The normalized accretion rates $\dot{m}$ (i.e., the Eddington ratio) were derived in \citetalias{Hagen_2024b} by fitting the HSC and eROSITA data with the model \textsc{agnsed}. Note hard X-ray dominates the bolometric luminosity in low accretion bins, which are not fully shown in Figure \ref{fig:result}. $N_{\rm H, int}$ is from the eFEDS AGN catalog, with unit of cm$^{-2}$.
}
\end{tablenotes}
\end{table*}
\end{center}

\subsection{GALEX}
\par The Galaxy Evolution Explorer \citep[GALEX,][]{Martin_2005} performed an all-sky survey in two ultraviolet bands: far-UV (FUV, 1350-1750\,\AA) and near-UV (NUV, 1750-2750\,\AA). To conduct an unbiased analysis of the UV properties for both luminous and faint sources, we take both UV detected and undetected sources into account. First we cross-match our sample with the GALEX legacy data release GR6plus7\footnote{\url{https://galex.stsci.edu/GR6/}}, includes 44884 imaging tiles \citep{Bianchi_2017}. 941 sources from the original sample are found to have a FOV offset $\leq$ 0.5$^\circ$ \citep{Bianchi_2014} in at least one effective tile with both FUV and NUV exposures $>$\,1\,s.   

\par We then search the GALEX\_GR6Plus7 catalog\footnote{\url{https://galex.stsci.edu/casjobs/default.aspx}} for GALEX counterparts of these sources, requiring a match radius of 2.6\arcsec \citep{Trammell_2007}, to get 
616 sources with GALEX counterparts. 
For sources detected in multiple tiles, we adopt the one with the longest FUV exposure time. 

\par We adopt a 3$\sigma$ criterion to decide if they are detected in each band, and calculate the 3$\sigma$ upper limits when not. Out of the 616 sources, 394, 509 and 373 sources are detected at 3$\sigma$ level in FUV, NUV and both bands, respectively. 

\par For sources without a GALEX counterpart, we estimate the 3$\sigma$ detection limits based on empirical relationships with exposure time as follows,  
\begin{equation}
    \rm mFUV_{lim} (mag) = 1.61 \times log10(t_{NUVexp}) + 18.36 
\end{equation}
\begin{equation}
     \rm mNUV_{lim} (mag) = 1.10 \times log10(t_{NUVexp}) + 19.94
\end{equation}
Note these revised relationships derived in \citet{Cai_2023} lead to a more conservative constraint of the undetected sources (larger upper limits), compared with the 50\%-complete detection limits derived in \citet{Vanden_2020}. 

\subsection{SDSS photometries}

\par The eFEDS field is covered by the Sloan Digital Sky Survey \citep[SDSS,][]{York_2000}, which provides the \textit{u}-band photometry complementary to HSC (\textit{grizy}). We search the archival SDSS photometric observations\footnote{\url{https://skyserver.sdss.org/CasJobs/}} \citep[Data Release 18,][]{Almeida_2023} for our sample, and successfully find SDSS counterparts for 1124 sources (matching radius = 1$\arcsec$). Following the official recommendation on the use of photometric processing flags, we select only data with 
``mode = 1'' and ``clean = 1'', leaving 1103 sources. For each source, we adopt the {\sc modelMag} and its error (\textit{u} and \textit{err\_u} in the SDSS catalogue), and convert them into flux and its error based on the asinh magnitude system \citep{Lupton_1999}, after performing a correction of \textit{u}-band zeropoint by 0.04 mag\footnote{\url{https://www.sdss4.org/dr17/algorithms/fluxcal/}}. Adopting a $3\sigma$ criterion, we obtain fluxes for 690 sources, and upper limits for 413 sources.

\subsection{KiDS}

\par The eFEDS field also partially overlaps with the KiDS-N field, part of the Kilo-Degree Survey \citep[][]{Jong_2013}, which provides \textit{u}-band photometry with a 5$\sigma$ limiting magnitude of $\sim$\,24.8, deeper than SDSS-\textit{u} ($\sim$\,22.15) by more then two magnitudes. We first crossmatch our sample with the \textit{u}-band observations table\footnote{\url{https://kids.strw.leidenuniv.nl/DR4/data_table.php}} of KiDS DR4 \citep{Kuijken_2019}. 756 sources are covered by at least one \textit{u}-band image with an effective exposure, i.e., located within the $1^\circ \times 1^\circ$ FOV. Searching the ESO Science Archive for these sources, we obtain 589 KiDS counterparts after dropping 133 sources with photometry flag ``uflag $>$ 0''. 

With a $3\sigma$ criterion we derive fluxes for 487 
and upper limits for 73 sources. 
For the 29 sources without KiDS counterparts, we adopt the $5\sigma$ limiting magnitude, (given in the KiDS observation table) as their upper limits.

\subsection{Galactic and intrinsic extinction} \label{subs:extinction}

\par We correct each source for Galactic extinction using the dust map of \citet{Schlegel_1998}, with the Python interface \textsc{dustmaps} \citep{Green_2018}. 
We assume a reddening law with $\frac{A_{\rm V}}{E(B-V)} \equiv R_{\rm V} = 3.1$ \citep{Fitzpatrick_1999}, and calculate $A_{\lambda}$ for each band using empirical Milky Way dust extinction coefficients $R_{\lambda}$, which are 6.783, 6.620 for FUV, NUV bands \citep{Yuan_2013}, and 4.239 for the \textit{u}-band \citep{Schlafly_2011}.

\par Intrinsic extinction from the host galaxy and/or nucleus is more challenging to constrain, and becomes more important at the shorter wavelengths included here than in the original study of \citetalias{Hagen_2024b}.
All the AGN were selected from eROSITA to have low column density $N_{\rm H, int}\leq 10^{22}$~cm$^{-2}$, derived using a Bayesian analysis \citep{Liu_2022}. 
These $N_{\rm H, int}$ typically can have large individual uncertainties, though most are consistent with zero as a lower limit. 
So it is possible that there is no cold gas associated with the host galaxy and/or AGN, giving a lower limit to the extinction correction. 
However, we conservatively assume that the best fit $N_{\rm H, int}$ is associated with dusty gas. We first assume that this is from the AGN environment, which likely has a significantly lower dust-to-gas ratio ($\sim$\,1\% -- 10\%) than the Galactic ISM \citep[e.g.,][]{Maiolino_2001, Esparza_2021, Jun_2021}. There are a variety of proposed AGN dust extinction laws \citep[e.g.,  ][]{Czerny_2004, Gaskell_2004, Gaskell_2007, Temple_2021}, but we 
use an SMC extinction 
curve \citep{Gordon_2003}\footnote{For sources with FUV data bluer than 1000\,\AA\ at rest frame, we linearly extrapolate the extinction curve.} as it gives more extinction in the UV than other AGN extinction curves \citep{Li_2007}. If we adopt a flatter extinction curve (less extinction in the UV), the estimated intrinsical UV luminosity would decrease, making the lack of UV emission in the low accretion bins even more significant.
We derive $E(B-V)$ for each individual object assuming
log $[\frac{E(B-V)}{N_{\rm H}}]= -22.8$ \citep{Jun_2021}. 
We then calculate the extinction-corrected fluxes for each band and each source. Different intrinsic extinction prescriptions are explored in Appendix \ref{Appendix:A}, but these do not make a large change in the derived spectra.

\subsection{Intrinsic Luminosity} \label{subs:lumi}

\par The extinction-corrected fluxes are then converted into rest-frame monochromatic luminosities $\nu L_{\nu}$ using the luminosity distances, $d_{\rm L}$, calculated from the source redshifts assuming cosmological parameters of Planck 2018 results \citep{Planck_2020}. The effective wavelengths of these bands at the observational frame are adopted as 1528, 2271 and 3543\,\AA, for FUV, NUV and \textit{u}-band respectively, which are converted to rest-frame wavelengths/frequencies according to the redshift of each source. We estimate the monochromatic luminosity at rest-frame 1200\,\AA, $L_{1200}$, for each GALEX source by interpolating FUV and NUV data supposing a power law SED. $\nu L_{1200}$ is treated as a detected value if the source is simultaneously detected in both FUV and NUV, otherwise as an upper limit. 

\par In \citetalias{Hagen_2024b}, this sample is further divided into 8 luminosity bins based on $\nu L_{3500}$, ranging from log\,$\nu L_{3500}$ = 42.1 to 45.3, to investigate how the SED changes with the luminosity/accretion rate. We tabulate the properties of each bin, together with the new SED points in 
Table~\ref{tab:result}. This shows that the AGN in the more
luminous bins have higher redshifts suggesting our flux-limited sample is likely incomplete.
This is initially surprising as the HSC optical selection is for host galaxies of similar masses, so similar luminosities, and the integrated (model) X-ray power does not change significantly between each bin \citepalias{Hagen_2024b}.
However, eROSITA only covers the $0.3 - 8.0$\,keV bandpass, which does change by $\sim$\,an order of magnitude, due to changes in the coronal photon index and the strength of the soft X-ray excess. Hence our sample could miss sources that appear faint in this energy range at higher redshifts.

\par Following \citetalias{Hagen_2024b}, we stack the UV data in each bin, by averaging the logarithmic luminosity log\,$\nu L_{\nu}$ and logarithmic rest-frame wavelengths for each band. We use survival analysis to assess the upper limits together with the detected fluxes, using \textsc{asurv} \citep{Feigelson_1985}. We use the Kaplan-Meier method to estimate the mean luminosity, along with the $1\sigma$ uncertainty of the mean by bootstrapping the sources in each bin, except for the faintest bin which has too few detections to give a robust result (see Table \ref{tab:result}).

\par We subtract the host galaxy contribution in the UV by 
fitting an Sb galaxy template spectrum \citep{Silva_1998} \citep[part of the SWIRE library:][]{Polletta_2007} to the HSC host galaxy optical phototometry (see Figure A1 in \citetalias{Hagen_2024b}).
We use this to calculate the rest-frame monochromatic luminosity of the host galaxy in each UV band for each source. We then derive the stacked intrinsic AGN luminosity in each bin by subtracting the mean luminosity of the corresponding host galaxies (Table \ref{tab:result}). 

\par We adopt a Sb template because the host galaxy in the majority of our sources is disc dominated rather than bulge dominated \citep[65\% have Sersic index $n < 2$,][]{Li_2024}, and we do see lots of spiral features in the HSC fitting residual images by visual check. We confirm the absence of strong intrinsic UV emission will remain significant for these faint bins even if adopting an UV-fainter template (e.g., S0). Meanwhile, the choice of galaxy templates will barely alter the intrinsic SED of the bright bins where the host galaxy is always negligible compared to the AGN in the UV.

\begin{figure*}
\centering
\subfloat{\includegraphics[width=0.99\textwidth]{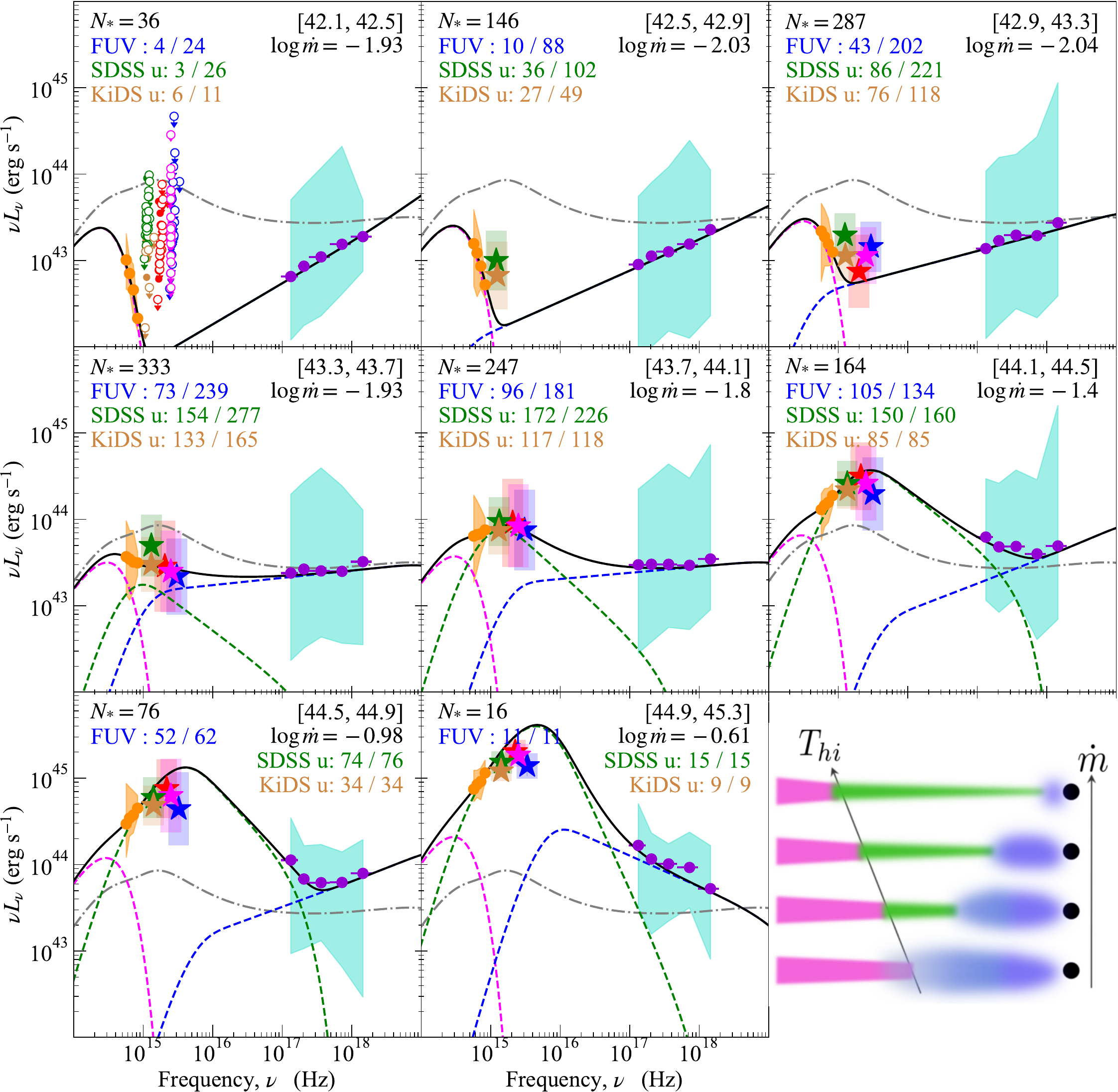}}
\caption{\label{fig:result}
Adapted from \citetalias{Hagen_2024b}. Each panel corresponds to a single 3500\,\AA\ luminosity bin (given in the top right corner), and shows the model SED along with the stacked data.
Stars show the UV data derived in this work, among which blue 
is for GALEX FUV, red is for GALEX NUV, magenta is for $\nu L_{1200}$ estimated by interpolating NUV and FUV 
data, green is for SDSS \textit{u}, and brown is for KiDS \textit{u}-band, respectively. Corresponding shadow regions show the 1$\sigma$ dispersion of the Kaplan-Meier distribution taking into account the censored data. The estimated intrinsic GALEX luminosities in the [42.5, 42.9] bin are $< 10^{42}$ $\rm erg\,s^{-1}$, outside the scope of this figure.
The orange and purple points mark the stacked HSC and eROSITA data respectively, with the orange and cyan regions indicating their dispersions (taken from \citetalias{Hagen_2024b}). 
The coloured dashed lines show the model components, which are a standard outer disc (magenta), warm Comptonising disc (green), and inner hot X-ray plasma (blue).
The dashed-dotted grey line shows the model of the middle bin to highlight the evolution of the SED with changing luminosity/$\dot{m}$. See \citetalias{Hagen_2024b} for a detailed description of the data, model and fitting process. The number of sources in each bin ($N_*$), the corresponding log\,$\nu L_{3500}$ range, mean $\dot{m}$, and the number of sources detected/covered by each band, are shown by the text in each bin. 
}
\end{figure*}  

\subsection{Variability} \label{subs:variability}

\par Our optical-to-X-ray SED consists of data from five different intruments spanning $\sim$ 20 years. Most SDSS photometries were conducted between 1998 and 2002, with a minority conducted in 2006; GALEX data were derived by coadding the data spanning 2003 to 2012; KiDS data were taken between 2011 and 2018, HSC photometries between 2014 and 2020, and eFEDS data in 2019. By stacking the SEDs of a large number of sources, we should have largely eliminated the effect of stochastic variations. Meanwhile, although dramatic ``changing-state'' variations can happen within years or decades \citep{Noda_2018}, such events should be too rare to bias our stacked SED. Furthermore, by comparing the $u$-band photometries of SDSS and KiDS in individual sources, we confirm our sample shows no systematic brightening or darkening between $\sim$ 2000 and 2015, and the stacked $u$-band data are also well consistent. We thus conclude the AGN variability should only contribute to some scatter of the stacked SED instead of biasing the shape.

\section{The SEDs including UV data}

Figure \ref{fig:result} shows the resulting \textit{u} and UV photometric data points added to the optical-X-ray SEDs of \citetalias{Hagen_2024b}. It is immediately evident that extending the data into the UV bandpass has confirmed the main conclusion of \citetalias{Hagen_2024b}: that the 
the faintest three bins with $\dot{m}$ $\lesssim $ 0.01 are intrinsically very faint in the UV, consistent with a transition of the optically thick, warm Comptonising disc
to a hot inefficient X-ray plasma.

The models are also in good agreement with the new data in the three moderate luminosity bins, following the predicted rise of the optically thick component. 
However, the three highest luminosity bins start to show a small deviation from the model predictions. The data turn down in the $\nu L_{\nu}$ SED between two GALEX bands, corresponding to a break at $\sim$\,800\,\AA\ -- 1400\,\AA. This 
has been widely observed before \citep{Zheng_1997, Telfer_2002, Shull_2012, Stevans_2014, Lusso_2015} but our sample clearly show that this is not an artefact of dust extinction as the $N_{\rm H, int}$ is lower in these bins (see Table \ref{tab:result}). This turnover appears intrinsic, indicating AGN with $\dot{m} \gtrsim 0.1$ produce less UV emission than the prediction of a warm Comptonised disc, or a standard disc which predicts an even bluer spectrum \citep[see e.g. Fig 1 of][]{Kubota_2018}.

The composite disc model used by 
\citetalias{Hagen_2024b} 
assumes the standard thin disc emissivity, but allows flexibility in how this power is emitted. It assumes a 
radially stratified structure, with an outer, standard blackbody region for $R>R_{\rm warm}$, a middle region where thermalisation is incomplete so the optically thick disc material instead emits a
warm, thermal Comptonisation spectrum, and then transitions to an inner hot, optically thin region for $R_{\rm hot}<R<R_{\rm isco}$ to power the X-ray corona \citep[see also][]{Kubota_2018}.
These three-region models can give a good match to the typical spectra seen in bright AGN, generally with 
$R_{\rm warm}$ systematically decreasing and $R_{\rm hot}$ systematically increasing with decreasing
$\dot{m}$ \citep{Kubota_2018,Mitchell_2023,Temple_2023b}. At lower luminosities, $\dot{m}\lesssim 0.01 $, $R_{\rm warm}$ meets $R_{\rm hot}$ 
so these low luminosity AGN only require two regions: an outer standard disc which truncates at large radius into the X-ray hot, radiatively inefficient inner flow. 

\begin{figure*}
    \centering
    \includegraphics[width=\textwidth]{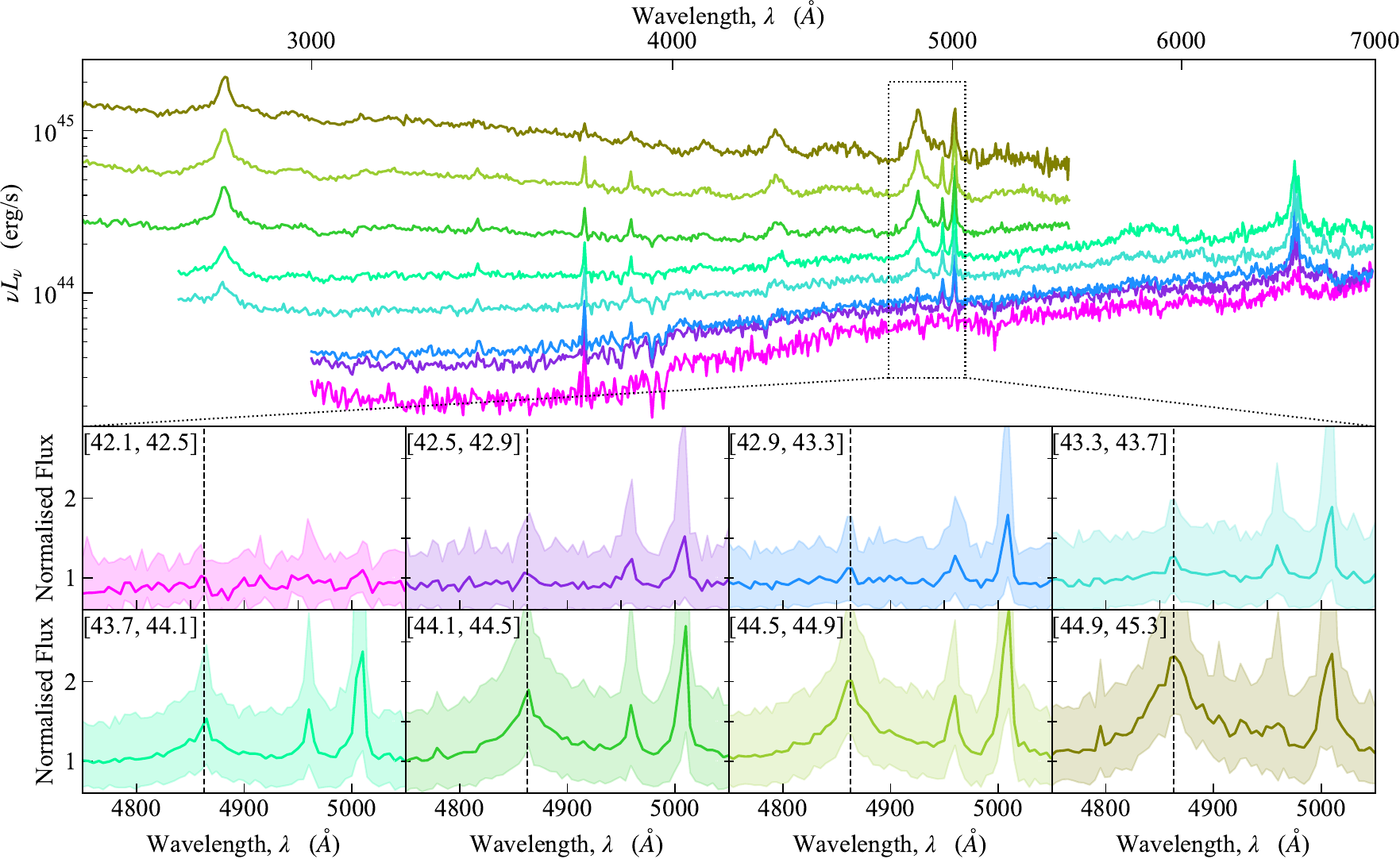}
    \caption{\textit{Top}: Composite SDSS spectra for each luminosity bin in our sample with a resolution of 5\AA.  The colours cycle from green (high luminosity) to magenta (low luminosity). The black dotted box indicates the region covering the H$\beta$ and [\ion{O}{III}] lines. \textit{Bottom}: Zoom-in on the H$\beta$-[\ion{O}{III}] region for each luminosity bin (given in the top left corner), normalised by $\nu L_{5100}$. Here the shaded region shows the $1\sigma$ dispersion in the stacks, and the solid line shows the mean spectrum. The vertical dashed black line indicates the H$\beta$ rest wavelength. It is clear that the broad emission lines disappear in the low luminosity bins, but remain prominent in the high luminosity bins.}
    \label{fig:sdssSepc}
\end{figure*}

While these continua match very well in general to the new data, they are continuum models, and predict a continuous increase in the blue/UV  extent of the warm Comptonisation region (see schematic at bottom right of Fig. \ref{fig:result}, taken from \citetalias{Hagen_2024b}). Hence they cannot match the downturn that appears in the data at fixed $\lambda\sim 1000$\,\AA\, (see the Appendix \ref{Appendix:C} for the model re-fitted including the UV data). This may instead indicate that atomic processes are also important in forming this feature, such as produced by e.g. a UV line driven wind, which can be efficiently produced for temperatures above 50,000\,K \citep{Laor_2014}. Alternatively it could be explained by different disc-based models, e.g., the magnetically-dominated accretion discs simulated by \citet{Hopkins_2024}, in which the effective temperature has a much weaker dependence on mass and mass-accretion rate. Testing these will be the subject of a future paper.

\section{Impact on changing UV ionising spectrum on the BLR}
 \label{sec:spec}
 
It is clear that the dramatic drop in the UV continuum at $\dot{m}$ $\lesssim$ 0.01 will give a similarly dramatic drop in the photoionising continuum.
This could naturally account for the absence of broad emission lines in low-luminosity AGN \citep{Ho_1999, Trump_2011}, as well as the disappearance of broad emission lines in the faint state of `changing-look' (more properly termed `changing-state') AGN.

We can now test this on our sample using the recently released
SDSS DR18 optical (BOSS) spectroscopic data for about 13000 eFEDS sources\footnote{\label{footnote:8}\url{https://www.sdss.org/dr18/bhm/programs/spiders/}}. Crossmatching the SDSS DR18 database to our sample, using a radius of 1$\arcsec$, we obtain spectra of 438 individual sources.
For sources with multiple spectra, we use the spectrum with the highest average signal-to-noise (S/N) ratio. We remove all spectra  with average S/N $<$ 1, which leads to 412 sources (around 33\% of our sample). The fraction of sources with SDSS spectra is higher in brighter bins, as expected due to the selection criteria of SDSS (flux cut of $i$ and $z$ band, see Footnote \ref{footnote:8} for details).

\par Each spectrum is dereddened for the Galactic extinction, using the dustmap of \citet{Schlegel_1998} and extinction curve of \citet{Cardelli_1989} with $R_{\rm V} = 3.1$, de-redshifted to rest frame, resampled onto a uniform wavelength grid with a resolution of 5\,\AA, and then converted to $\nu L_{\nu}$. We obtain the composite spectrum for each luminosity bin by averaging the logarithmic $\nu L_{\nu}$ at each wavelength, shown in the top panel of Figure \ref{fig:sdssSepc}. The effective wavelength range is slightly different for each 3500\,\AA\ luminosity bin due to their different redshift distributions. 

It is clear that the optical/near UV spectrum changes with increasing AGN optical luminosity, from 
red and galaxy-dominated to a typical blue AGN spectrum. The BLR clearly follows this change, 
with no broad emission lines in the faintest three bins, and then increasing BLR strength as the blue/UV continuum increases above 
$\dot{m}\gtrsim 0.01$. The bottom, smaller, panels in Figure \ref{fig:sdssSepc} highlights this evolution, showing a zoom in around the  H$\beta$-[\ion{O}{III}] lines.  

\par
\begin{figure}
\centering
\subfloat{\includegraphics[width=0.48\textwidth]{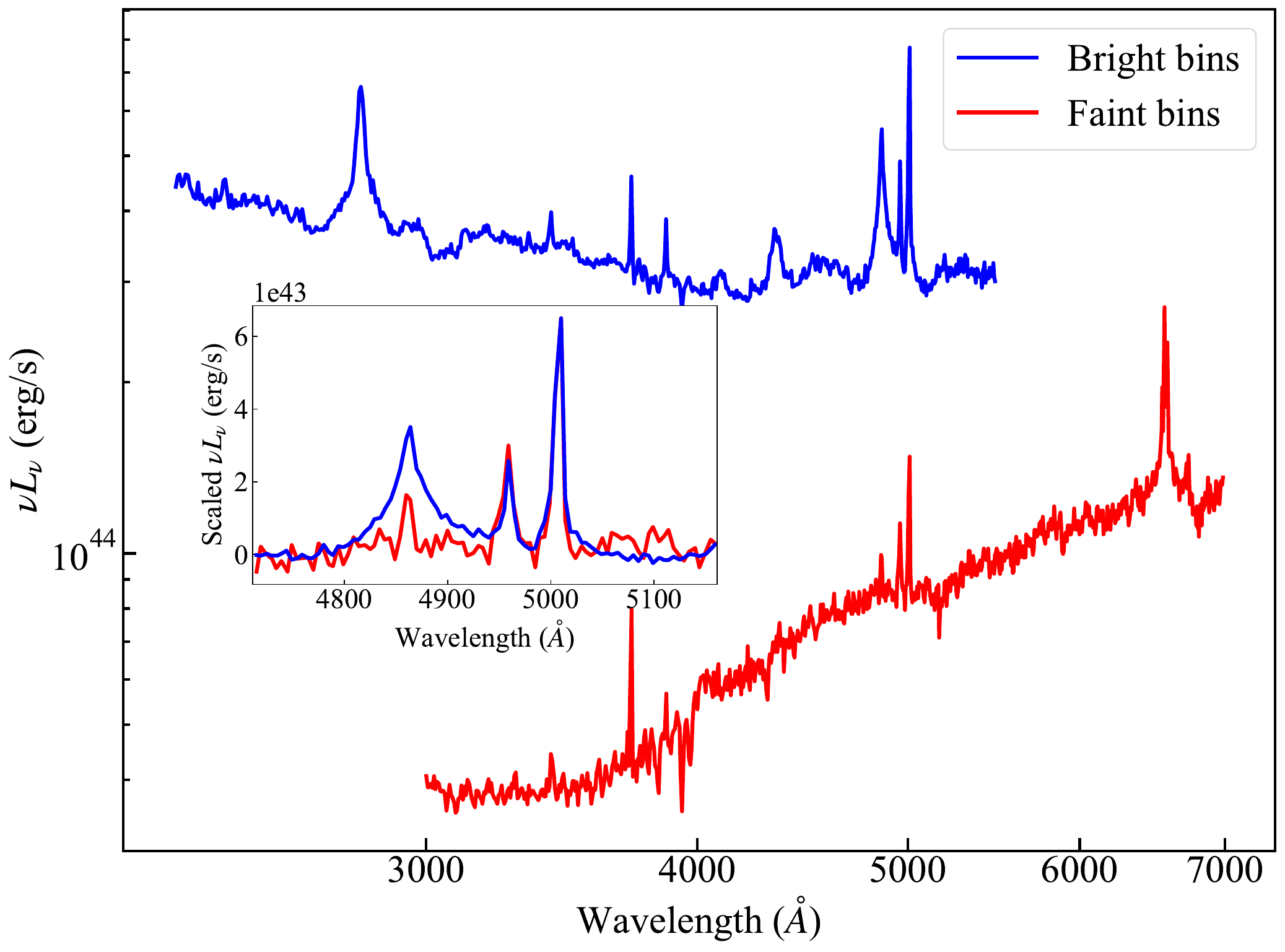}}
\caption{\label{fig:two_spectra} The composite SDSS spectra for sources in the faintest three bins and brightest three bins respectively. Inset: the H$\beta$-[\ion{O}{III}] profiles after subtracting a constant continuum, with the spectrum of the bright bins scaled to match  the [\ion{O}{III}] 5007\,\AA\ luminoisity.
}
\end{figure}

 Figure \ref{fig:two_spectra} shows this even more clearly from a composite spectrum of the 
the lowest three luminosity bins (log\,$\nu L_{3500}$ in [42.1, 43.3]) compared to that of the last three bins (log\,$\nu L_{3500}$ in [44.1, 45.3]). We subtract a constant continuum component to illustrate the profiles of the lines \citep[e.g.,][]{Mahmoud_2020}, as shown in the inset of Figure \ref{fig:two_spectra}. These spectra are normalised to match the narrow [\ion{O}{III}] luminosity, showing a clear change in the H$\beta$ line. The broad component dominates at high luminosities, but it is absent or extremely weak for sources with $\dot{m}$ $\lesssim $ 0.01\footnote{For a possible alternative scenario, HST narrow slit spectroscopy has revealed extremely broad emission lines in a low acrretion (log $\dot{m} \sim -4$) AGN NGC 3147\citep{Bianchi_2019, Bianchi_2022}. SDSS spectra cannot distinguish them.}. This gives further evidence that there is an intrinsic change in UV continuum for sources with $\dot{m}$ $\lesssim $ 0.01. We stress again that these all have low column densities so the BLR cannot simply be obscured.

\section{Discussion and Conclusions}

We have extended the sequence of optical-X-ray SEDs of AGN of \citetalias{Hagen_2024b}. The resulting stacked optical/X-ray SEDs for black holes at fixed mass 
show a dramatic transition, similar to that seen in stellar mass black hole binaries, where the
dominating disc component in bright AGN collapses into an inefficient X-ray plasma below $L/L_{\rm Edd} \sim 0.01$. The models fit to these datasets in \citetalias{Hagen_2024b} predicted the largest change in SED should be in the rest frame UV, which was not covered by their data, so we critically test this by  adding archival photometric data from SDSS-\textit{u}, KiDS-\textit{u} and GALEX-NUV and GALEX-FUV. We subtract a standard Sb host galaxy spectrum from these, and correct the AGN luminosity for SMC-like dust extinction assuming the Baysian best fit $N_{\rm H}$ derived from the X-ray spectral fits of \citet{Liu_2022}. The resulting stacked spectra confirm that there is 
indeed an abrupt transition of the accretion structure, as claimed by \citetalias{Hagen_2024b}, very similar to that observed in stellar mass black holes and a few individual `changing-look' AGN. 

Above the transition, the UV bright disc becomes increasingly dominant, though its shape starts to slightly deviate away from the models below $\sim1000$\,\AA\ for $L/L_{\rm Edd}\gtrsim 0.1 $, perhaps indicating the presence of UV line driven disc winds. 

Additionally, we explore the effect of this transition on the BLR. The recent release of SDSS DR18 includes many more spectra of the AGN in this sample than in the DR16 used in \citetalias{Hagen_2024b}, allowing us to make composite optical/near UV spectra for every AGN optical luminosity bin. 
The broad H$\beta$ line clearly tracks the blue continuum emission, dropping in intensity along with the continuum, and eventually disappearing below $\dot{m} \lesssim 0.01$ when the UV emitting warm disc also disappears. 

This is all strong confirmation that there is indeed a transition in the accretion flow in AGN, 
similar to that seen in stellar mass binaries. 
There is a real, and quite abrupt change in the accretion structure, driven by the changing Eddington ratio which leads to the complete loss of the ionising EUV emission and hence the characteristic broad emission lines. This has a significant impact on how we identify AGN through cosmic time, predicting that there are true Type 2 AGN, where the characteristic Type 1 AGN identifiers of strong BLR and UV continuum are intrinsically not present rather than obscured.

\section*{Acknowledgements}

\par This work is based on data from eROSITA, the soft X-ray instrument aboard SRG, a joint Russian-German science mission supported by the Russian Space Agency (Roskosmos), in the interests of the Russian Academy of Sciences represented by its Space Research Institute (IKI), and the Deutsches Zentrum für Luft- und Raumfahrt (DLR). The SRG spacecraft was built by Lavochkin Association (NPOL) and its subcontractors, and is operated by NPOL with support from the Max Planck Institute for Extraterrestrial Physics (MPE). The development and construction of the eROSITA X-ray instrument was led by MPE, with contributions from the Dr. Karl Remeis Observatory Bamberg \& ECAP (FAU Erlangen-Nuernberg), the University of Hamburg Observatory, the Leibniz Institute for Astrophysics Potsdam (AIP), and the Institute for Astronomy and Astrophysics of the University of Tübingen, with the support of DLR and the Max Planck Society. The Argelander Institute for Astronomy of the University of Bonn and the Ludwig Maximilians Universität Munich also participated in the science preparation for eROSITA. The authors gratefully acknowledge Andrea Merloni's contribution to the eROSITA project in general and this sample in particular. 

\par Based on observations made with ESO Telescopes at the La Silla Paranal Observatory under programme IDs 177.A-3016, 177.A-3017, 177.A-3018 and 179.A-2004, and on data products produced by the KiDS consortium. The KiDS production team acknowledges support from: Deutsche Forschungsgemeinschaft, ERC, NOVA and NWO-M grants; Target; the University of Padova, and the University Federico II (Naples). This research has made use of the VizieR catalogue access tool, CDS, Strasbourg, France.

\par JK gratefully acknowledges the scholarship of University of Science and Technology of China for a visiting program to Durham University, and thanks Zhen-Yi Cai for his help with the GALEX database. 

\par SH acknowledges support from the Japan Society for the Promotion of Science (JSPS) through the short-term fellowship PE23722 and from the Science and Technologies Facilities Council (STFC) through the studentship grant ST/W507428/1. CD acknowledges support from STFC through grant ST/T000244/1 and Kavli IPMU, University of Tokyo. Kavli IPMU was established by World Premier International Research Center Initiative (WPI), MEXT, Japan. JS is supported by JSPS KAKENHI (23K22533) and the World Premier International Research Center Initiative (WPI), MEXT, Japan. This work was supported by JSPS Core-to-Core Program (grant number: JPJSCCA20210003). MJT is supported by STFC through grant ST/X001075/1.

\section*{Data Availability}
\par The optical and X-ray data as well as the SEDs are from \citet{Hagen_2024b}. The GALEX data are available at \url{https://galex.stsci.edu/casjobs/default.aspx}, SDSS data at \url{https://skyserver.sdss.org/CasJobs/}, and KiDS data at \url{https://www.eso.org/qi/catalogQuery/index/260}. The data in this paper will be shared on reasonable request to the corresponding author.

\appendix

\section{The impact of different intrinsic extinction prescriptions} \label{Appendix:A}

Reconstructing the UV spectrum is dependent on the intrinsic extinction assumed, so here we assess the impact of different methods, focussing only on the new wavelength range used in this work. 

\begin{figure*}
\centering
\subfloat{\includegraphics[width=0.99\textwidth]{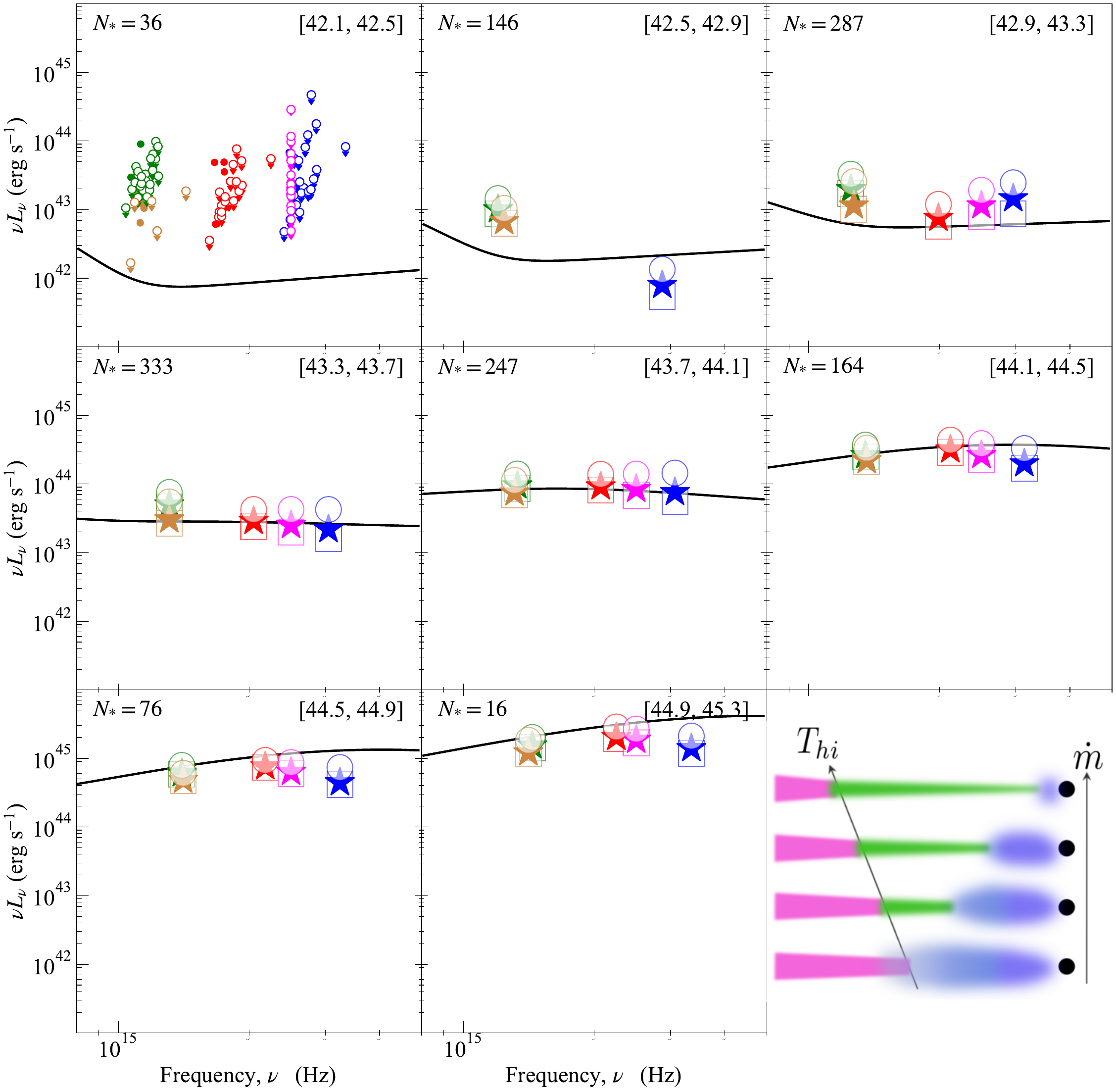}}
\caption{\label{fig:appendix} The UV SEDs when assuming different intrinsic extinction. Squares: no intrinsic absorption. Stars: all the observed $N_{\rm H}$ is associated with AGN absorption (SMC-like dust, as adopted in this work). Circles: all the observed $N_{\rm H}$ is due to the host galaxy (Milky Way-like dust).
}
\end{figure*}

The Baysian X-ray spectral fitting for the eROSITA data gives best fit and error range for the neutral gas column, and this error range is generally consistent with zero column. A minimal extinction correction is then if there is no  intrinsic absorption present. This gives the lower set of data points (squares) in Figure \ref{fig:appendix} for the derived UV SEDs. Instead, in the main paper, we use the best fit intrinsic X-ray column density for each object, assuming that the column is associated with the AGN torus, so has large grains similar to 
an SMC extinction curve giving low dust-to-gas ratio (see \S \ref{subs:extinction} for details), giving the middle set of data points (stars) in Figure \ref{fig:appendix}. Finally, we assume a maximal correction by assuming the best fit $N_{\rm H}$ is due to the host galaxy (circles), so has Milky Way extinction, with higher dust-to-gas 
ratio $N_{\rm H} (\rm cm^{-2}) = 2.21 \times 10^{21}$ $ A_{\rm V} (\rm mag)$ \citep{Guver_2009}, with $R_{\rm V} = A_{\rm V}/ E(B-V)= 3.1$ \citep{Cardelli_1989}. 

The results show that the UV SED shape is not substantially altered by these assumptions. In particular, the main conclusion clearly remains, that the UV is intrinsically weak in the lowest 3 luminosity bins, consistent with the models from \citetalias{Hagen_2024b} (black line) where the warm Comptonising disc (green region in the schematic to the lower right) disappears completely, so that most of the accretion power is emitted in the hot plasma (blue region). 
It is also clear that the highest luminosity bins still have UV
spectra which bend away from the continuum models below $\sim 1000$\AA.

\section{The intrinsic distribution of $N_{\rm H}$ in each luminosity bin} \label{Appendix:B}

\par
\begin{figure}
\centering
\subfloat{\includegraphics[width=0.49\textwidth]{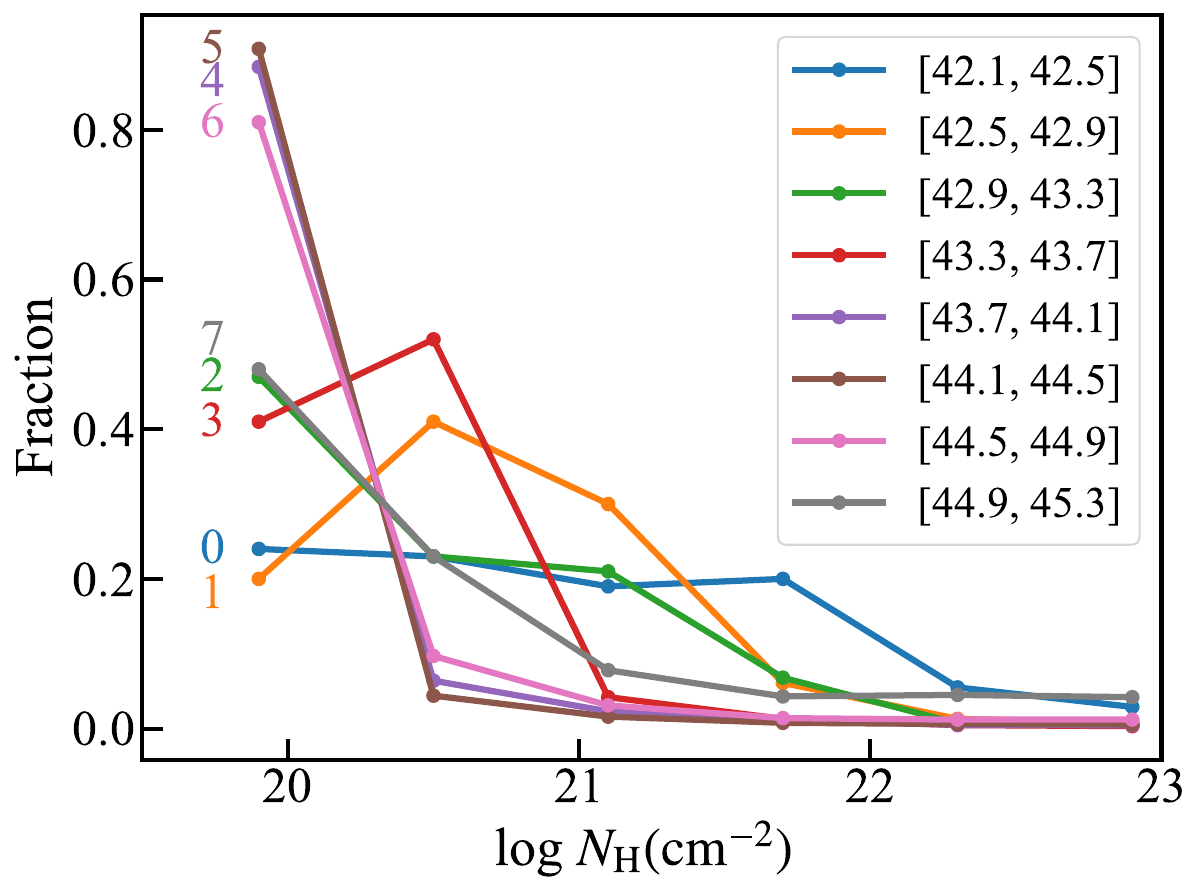}}\\
\subfloat{\includegraphics[width=0.48\textwidth]{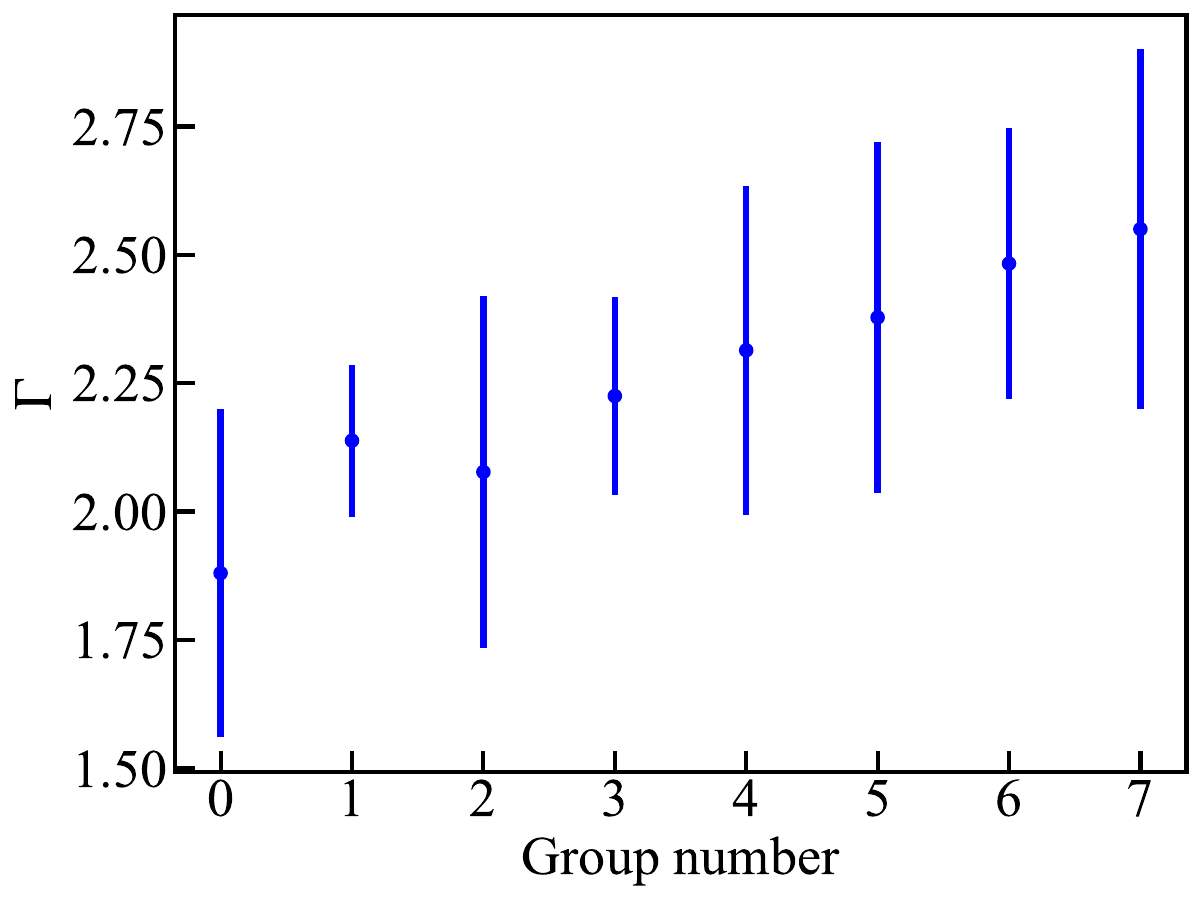}}\\
\caption{\label{fig:NH_and_Gamma} The intrinsic distribution of $N_{\rm H}$ (upper panel) and $\Gamma$ (lower panel) in each luminosity bin, derived by Hierarchical Bayesian modeling.
}
\end{figure}

 In this work we have utilised the $N_{\rm H, int}$ obtained from the eFEDS catalogue. To further confirm its reliability, we use the Hierarchical Bayesian modeling to directly recover the intrinsic distribution of $N_{\rm H, int}$ in each luminosity bin by supposing a simple absorbed power law \citep[see][for details]{Liu_2022}. As shown in Figure \ref{fig:NH_and_Gamma}, although there could be some moderately absorbed ($N_{\rm H, int} > 10^{21} \rm cm^{-2}$) sources in those faint bins (number 0,1 and 2), the majority of the sources are absorption-free. Meanwhile, the intrinsic distribution of the photon index $\Gamma$ is clearly correlated with the luminosity (softer X-ray spectrum in brighter bin), consistent with the stacked X-ray spectra, though the 
 absolute value of $\Gamma$ is different as no soft excess component is included in Bayesian modeling.
 
 These together indicate that the evolution of the observed SED as well as the lack of broad emission lines in the faint sources, cannot be attributed to contamination of heavily absorbed/reddened sources in the low luminosity bins. 

\section{Fitting the full-band SED including the UV data} \label{Appendix:C}
\begin{figure*}
\centering
\subfloat{\includegraphics[width=0.99\textwidth]{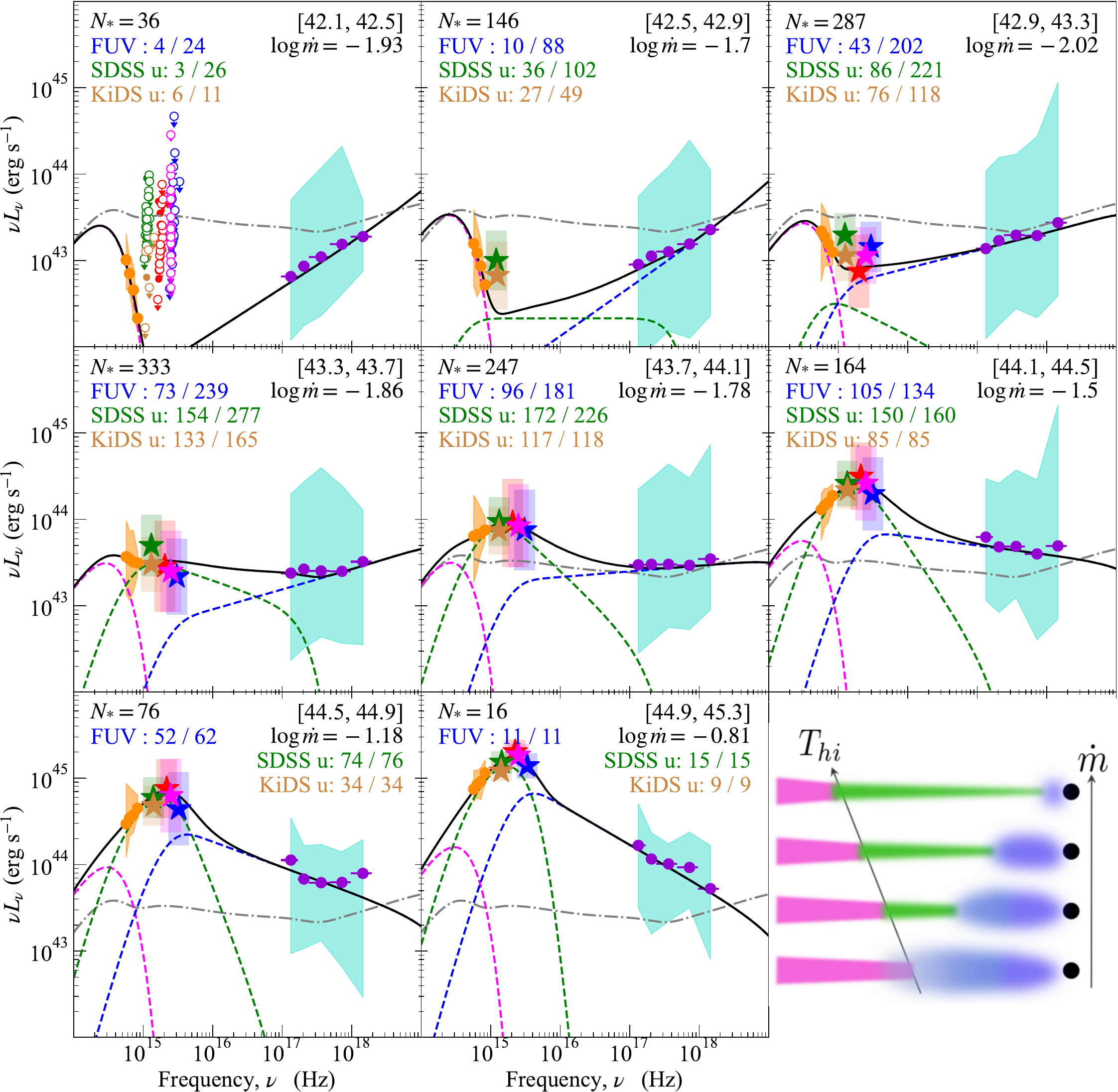}}\\
\caption{\label{fig:fullband} Same as Figure \ref{fig:result}, but the SED models are re-fitted including the UV data.  
}
\end{figure*}

Here we present the best-fit models by fitting the full-band SED including the new UV data in Figure \ref{fig:fullband}, and the main results stay the same. Some contribution from the warm corona is now needed in the second and third faintest bins due to a tiny excess of the UV data, though we stress large uncertainties exist in these photometries induced by the estimation of host contamination and intrinsic extinction. Meanwhile, for the three brightest bins, the warm corona becomes weaker and the acrretion rate slightly drops, due to the FUV break. Trying to fit this break also leads to fitting residuals in the X-ray band, especially in the second and third brightest bins.

\bibliographystyle{mnras}
\bibliography{example} 

\bsp	
\label{lastpage}
\end{document}